\documentclass[preprintnumbers,amsmath,amssymb,floatfix,11pt,onecolumn,prd,superscriptaddress,nofootinbib,showpacs,showkeys]{revtex4}
\usepackage{graphicx}
\usepackage{epsfig}
\usepackage{bm}
\usepackage{amsfonts}

\def\ben{\begin{equation}}
\def\een{\end{equation}}

 \def\bd{\begin{document}} \def\ed{\end{document}}
\def\ds{\documentstyle} \let\fr=\frac \let\bl=\bigl \let\br=\bigr
\let\Br=\Bigr \let\Bl=\Bigl
\let\bm=\bibitem
\let\na=\nabla
\let\pa=\partial \let\ov=\overline
\newcommand{\be}{\begin{equation}}
\newcommand{\ee}{\end{equation}}
\def\ba{\begin{array}}
\def\ea{\end{array}}
\def\ft#1#2{{\textstyle{\frac{\scriptstyle #1}{\scriptstyle #2} } }}
\def\fft#1#2{{\frac{#1}{#2}}}
\def\del{\partial}
\def\vp{\varphi}
\def\sst#1{{\scriptscriptstyle #1}}
\def\oneone{\rlap 1\mkern4mu{\rm l}}
\def\td{\tilde}
\def\wtd{\widetilde}
\def\ie{{\it i.e.\ }}
\def\dalemb#1#2{{\vbox{\hrule height .#2pt
        \hbox{\vrule width.#2pt height#1pt \kern#1pt
                \vrule width.#2pt}
        \hrule height.#2pt}}}
\def\square{\mathord{\dalemb{6.8}{7}\hbox{\hskip1pt}}}
\newcommand{\ho}[1]{$\, ^{#1}$}
\newcommand{\hoch}[1]{$\, ^{#1}$}
\newcommand{\bea}{\setlength\arraycolsep{2pt} \begin{eqnarray}}
\newcommand{\eea}{\end{eqnarray}}
\newcommand{\ra}{\rightarrow}
\newcommand{\lra}{\longrightarrow}
\newcommand{\Lra}{\Leftrightarrow}
\newcommand{\bp}{\tilde \beta^\prime}
\newcommand{\tr}{{\rm tr} }
\newcommand{\Tr}{{\rm Tr} }
\def\0{{\sst{(0)}}}
\def\1{{\sst{(1)}}}
\def\2{{\sst{(2)}}}
\def\3{{\sst{(3)}}}
\def\4{{\sst{(4)}}}
\def\5{{\sst{(5)}}}
\def\6{{\sst{(6)}}}
\def\7{{\sst{(7)}}}
\def\8{{\sst{(8)}}}
\def\m{{\sst{(m)}}}
\def\n{{\sst{(n)}}}
\def\cA{{{\cal A}}}
\def\cB{{{\cal B}}}
\def\cF{{{\cal F}}}
\def\cG{{{\cal G}}}
\def\cH{{{\cal H}}}
\def\tV{\widetilde V}
\def\tW{\widetilde W}
\def\tH{\widetilde H}
\def\tE{\widetilde E}
\def\tF{\widetilde F}
\def\tA{\widetilde A}
\def\im{{{\rm i}}}
\def\tY{{{\wtd Y}}}
\def\ep{{\epsilon}}
\def\vep{{\varepsilon}}
\def\bD{{{\bar D}}}
\def\R{{{\mathbb R}}}
\def\C{{{\mathbb C}}}
\def\H{{{\mathbb H}}}
\def\CP{{{\mathbb C}{\mathbb P}}}
\def\RP{{{\mathbb R}{\mathbb P}}}
\def\Z{{{\mathbb Z}}}
\def\bA{{{\mathbb A}}}
\def\bB{{{\mathbb B}}}
\def\bC{{{\mathbb C}}}
\def\bD{{{\mathbb D}}}
\def\bE{{{\mathbb E}}}
\def\bZ{{{\mathbb Z}}}
\def\Re{{{\frak{Re}}}}
\def\Im{{{\frak{Im}}}}
\def\cosec{{\,\hbox{cosec}\,}}
\def\Gm{{\Gamma_{\!\! -}}}
\def\Gp{{\Gamma_{\!\! +}}}
\def\stan{{standard }}
\def\nonstan{{supernumerary }}
\def\p{{\partial}}
\def\kdel#1{{\fft{\del}{\del#1}}}

\def\bog{{Bogomolny }}
\def\om{{\omega}}

\newcommand{\nnr}{\nonumber \\}
\newcommand{\pd}{\partial}
\newcommand{\ud}{\textrm{d}}
\newcommand{\dTH}{T^{\prime \, 0}_\textrm{H}}
\newcommand{\dOi}{\Omega^{\prime \, 0}_i}
\newcommand{\bx}{{\bf x}}
\begin{document}

\title{ {\bf Holographic superconductors in the AdS
black hole with a magnetic charge}}
\author{\textbf{M. R. Setare}}
\email{rezakord@ipm.ir}
 \affiliation{Department of Science, Payame Noor University, Bijar, Iran }

\author{\textbf{D. Momeni}}
\email{d.momeni@yahoo.com}
 \affiliation{Eurasian International Center for Theoretical Physics,
 Eurasian National University, Astana 010008, Kazakhstan}

\author{\textbf{R. Myrzakulov}}
\email{rmyrzakulov@csufresno.edu}
 \affiliation{Eurasian International Center for Theoretical Physics,  Eurasian National University, Astana 010008, Kazakhstan}

\author{\textbf{Muhammad Raza,}}
\email{mreza06@gmail.com}
 \affiliation{Department of Computer Science, COMSATS Institute of
Information Technology (CIIT), Sahiwal campus, Pakistan}

\begin{abstract}
In this work we study the analytical properties of a $2+1$
dimensional magnetically charged holographic superconductor in
$AdS_4$. We obtain the critical chemical potential $\mu_c$
analytically, using the Sturm-Liouville variational approach.
Also, the obtained analytic result
can be used to back up the numerical computations in the holographic
superconductor in the probe limit.
\end{abstract}
\pacs{ 11.25.Tq}
 \keywords{ High-$T_C$ superconductors theory} 
 \maketitle
\section{Introduction}
In this paper, we will investigate the holographic superconductors
in magnetically charged planar AdS black hole. The metric of the
black hole with a magnetic charge was obtained by Romans \cite{ro}.
The presence of the magnetic charge shows that the black hole has
different horizon structure from that of the uncharged Schwarzschild black
hole.  The main purpose in this paper is to see how the magnetic
charge  affects the holographic superconductors in this asymptotic
AdS black hole. This paper is organized as follows: In Sec. II, we present the
metric describing a magnetically charged planar Schwarzschild-AdS
black hole. In Sec. III, we give the basic equations.  In Sec. IV we
investigate the zero temperature limit and critical chemical
potential. We study analytically holographic superconductors in the
magnetically charged planar black hole background. Our results show
that the magnetic charge presents the different physical effects on the
different condensations.
Finally, in the last section we present our
conclusions.
\section{Geometry of a magnetically charged black hole in $AdS_4$}
 The Lagrangian that gives rise to the magnetically charged black hole is
described by the common Lagrangian of Einstein-Maxwell theory with cosmological constant $\Lambda$ is given by \cite{ro}
\begin{equation}
\mathcal{L}=\frac{1}{q}\Big(-\frac{1}{4}R+\frac{1}{4}F_{\mu\nu}F^{\mu\nu}+\frac{1}{2}\Lambda\Big).
\end{equation}
Where $q$ is the electrical charge. The usual Einstein-Maxwell equations read
\begin{equation}
R_{\mu\nu}=T_{\mu\nu},\ \ T_{\mu\nu}=2F_{\mu\alpha}F^{\alpha}_{\nu}-\frac{1}{2}g_{\mu\nu}F_{\alpha\beta}F^{\alpha\beta},\ \
\nabla_{;\mu}F^{\mu\nu}=0.\label{eom}
\end{equation}
The  components of the  Maxwell tensor are
\be
F_{01}=\frac{Q}{r^2},\ \ F_{23}=-H \cos\theta.
\ee
We assume that the electric charge $Q$ and the magnetic charge $H$ simultaneously are not zero. The magnetically
charged planar solution for action (1) is \cite{ro}
\begin{equation}
ds^2=-f(r)dt^2+\frac{dr^2}{f(r)}+r^2(dx^2+dy^2).
\end{equation}
An exact solution of the equations (\ref{eom})  is
$$
f(r)=r^2-\frac{M}{r}+\frac{H^2}{r^2}.
$$
 This solution describes
a planar Schwarzschild-AdS black hole with a magnetic charge. As the magnetic charge
$H$ tends to zero, the space time reduces to a four
dimensional Schwarzschild-AdS black hole.
The only unique non vanishing 
 component $F_{xy}=\frac{H}{r^2}$.
 Thus the factor $H$ can be interpreted as the strength of the magnetic field  in bulk.
 It is obvious that this solution is
asymptotically $AdS_4$. The outer horizon locates at $f(h)=0$.
For future numerical purposes, we take the horizon as $h=1$. Thus
we have $M=1+H^2$.  Further, as we know , to have a
real horizon we must satisfy the auxiliary condition
$$
27 M^4-256
H^6\geq 0
$$
 We limit ourselves only to the extremal black hole,
i.e., such values of the magnetic charge satisfying
$$
27 M^4-256H^6=0,
$$
 which gives to us the value of
 $$H=1.73205$$.
  This is the
extremal magnetic field which leads to an extremal black hole.
The black hole Hawking-Unruh temperature reads as
$$
T_{BH}=\frac{f'(h)}{4\pi}.
$$
 This temperature can be read as the boundary temperature in the quantum dual theory, via CFT.
 We mention here that the near horizon geometry of the metric, at zero temperature has a
scaling invariance. This scaling invariance characterized by a dynamical critical exponent $z$.
\section{The condensate of charged operators}
In order to investigate the holographic superconductors in the
background of the planar Schwarzschild-AdS black hole with a
magnetic charge, we need  a scalar condensate with a charged scalar
field $\psi$ \cite{GT}. Here we work in the probe approximation by neglecting the
backreaction of the charged scalar field $\psi$.
It may be an interesting topic to study the holographic
superconductors of the scalar field with magnetic charge. However,
In this paper we only consider the condensate of an external charged
scalar field in the background of a black hole with a magnetic.
Let us consider a Maxwell field and a charged
complex scalar field,
\begin{equation}
S=\int
\sqrt{-g}d^4x[\frac{1}{q}\Big(-\frac{1}{4}R+\frac{1}{4}F_{\mu\nu}F^{\mu\nu}+\frac{1}{2}\Lambda\Big)-\mid
D_\mu \psi|^2-m^2|\psi|^2].
\end{equation}
We set the AdS radius $L=\sqrt{\frac{3}{\Lambda}}\equiv 1$. The mass of the scalar field is chosen such that it
remains below to the Breitenlohner-Freedman bound \cite{BF}.
 We take
\begin{eqnarray}
A_{\mu}=(\phi(r),0,0,0),\;\;\;\;\psi=\psi(r),
\end{eqnarray}
we can obtain the equations of motion for the complex scalar field
$\psi$ and electrical scalar potential $\phi(r)$ in the background
of the Schwarzschild-AdS black hole with a magnetic charge $H$
given by \cite{prd2008}
\begin{eqnarray}
\psi''+(\frac{f'}{f}+\frac{2}{r})\psi'+(\frac{\phi^2}{f^2}+\frac{2}{f})\psi=0\label{psi}\\,
\phi''+\frac{2\phi'}{r}=\frac{2\psi^2\phi}{f}\label{phi}.
\end{eqnarray}
Here a prime denotes the derivative with respect to
$r$. Trivially, it is not possible to obtain the nontrivial
analytical solutions to the nonlinear equations.
\section{Zero temperature limit and critical chemical potential}
  We analyze the field
equations using the variational  method . On horizon
the boundary condition $\phi(h)=\mu$  is the critical chemical
potential reads from the value of the field on boundary (horizon),
and for the scalar field $\psi'(h)=\upsilon(h,H)\psi(h)$ in which
the function $\upsilon(h,H)$ is a function of $h,H$. Asymptotic
value of the field $\psi$ leads to two distinct conformal dimensions
$\Delta_{\pm}={1,2}$. The first step for investigating the analytical properties of the
superconductors via variational method is in writing the field
equation in a classical Sturm-Liouville (S-L) form. We present our
calculations separately for different conformal dimensions. We
rewrite the field equations (\ref{psi}) and (\ref{phi}) in a new
coordinate $z$ as
\begin{eqnarray}
\psi''+\frac{f'}{f}\psi'+(\frac{h^2\phi^2}{z^4f^2}+\frac{2h^2}{z^4f})\psi&=&0\\
\phi''-\frac{2h^2\psi^2}{fz^4}\phi&=&0.
\end{eqnarray}
now prime denotes the derivative with respect to the $z$.
In these coordinates $z\rightarrow0$ will be the $AdS_4$ conformal boundary.
We must first find a solution to the above equations such that near the boundary $\phi\approx \frac{<O>}{z^2}$
is purely normalizable. Away from the boundary, the solution should end at a regular black hole horizon at $r = h$. Beside
the critical point $T=T_c, \mu=\mu_c$ we have $\phi=\mu(1-z)$,
also since in this point
$$
\psi\sim <O>z^{\Delta}F(z),
$$
 which in
that the new function $F(z)$ satisfies the following
Sturm-Liouville equation
\begin{eqnarray}
\frac{d}{dz}(p(z) F')+q(z)F=-\mu^2 w(z) F,
\end{eqnarray}
here
\begin{eqnarray}
f=\frac{h^4-h M z^3+H^2 z^4}{h^2 z^2}\\
p(z)=z^{2\Delta}f\\
q(z)=\Delta(\Delta-1)z^{2\Delta-2}f+\Delta z^{2\Delta-1} f'+2
h^2z^{2\Delta-4}\\
w(z)=\frac{h^2z^{2\Delta-4}(1-z)^2}{f}.
\end{eqnarray}
It's easy to show that the minimum value of the eigenvalue $\mu^2$
 which gives the minimum of the magnetic field H can be obtained from the following functional
\begin{eqnarray}
\mu^2=Min\{-\frac{\int_{0}^{1}(p(z)F'^2-q(z)F^2)dz}{\int_{0}^{1}w(z)F^2dz}\}\label{mu}.
\end{eqnarray}
The boundary condition is $p(z)F(z)F'(z)|_{0}^{1}=0$. From the limit of this boundary term in $z=0$ we observe
\begin{eqnarray}
\lim _{z\rightarrow0} p(z) F(z)F'(z)=0.
\end{eqnarray}
Using eq. (\ref{mu}) to compute the minimum eigenvalue of $\mu^2$ for $\Delta_i$ , $i = +$ or $i = -$, we can
obtain the critical temperature $T_c$. We have to consider $F(0)=0$ or $F'(0)=0$. The trial function
$F(z)=1-\alpha z^2$ is a good choice
 satisfying this restricted boundary condition $F'(0)=0$, $F(0)=1$. We analyze two cases $\Delta_{\pm}={1,2}$ separately.
\subsection{Case $\Delta=2$}
In this case   we have
\begin{eqnarray}
\mu^2(\alpha)=-\frac{1.333\alpha^2-0.1904\alpha+0.4}{0.0289\alpha^2-0.0873\alpha+0.0858}.
\end{eqnarray}
The minimum  with respect to the $\alpha$ locates at
\begin{eqnarray}
\alpha_c=-1.75697,
\end{eqnarray}
which leads to the following values for $\mu_c$
\begin{eqnarray}
\mu_c=1.86784,
\end{eqnarray}
we show that the $\mu_c$ obtained from variational method is in good agreement with
the results of the \cite{prd2008}.
\subsection{Case $\Delta=1$}
In this case,  we have
\begin{eqnarray}
\mu^2(\alpha)=-\frac{0.190476\alpha^2+0.4\alpha}{0.101774\alpha^2-0.171785\alpha+0.43521}.
\end{eqnarray}
The minimum  with respect to the $\alpha$ locates at
\begin{eqnarray}
\alpha_c=-0.780319,
\end{eqnarray}
which leads to the following values for $\mu_c$
\begin{eqnarray}
\mu_c=0.55744.
\end{eqnarray}
Referring to the numerical results obtained in \cite{prd2008} shows
that the $\mu_c$ obtained is in good agreement with the
results. The consistency between the analytic and numerical results indicates that the S-L method
is a powerful analytic way to investigate the holographic superconductor even when we
take the backreaction into account\cite{jhep}.
\section{conclusion}
 In this paper 
using an variational calculation we  obtained the minimum of the chemical potential $\mu_c$ . Our analytic result can be used to back up the numerical computations in the
holographic superconductor in the probe limit.
\section{comment}
The model and the method which we used here are completely different from the published paper \cite{NPB} in which they applied numerical methods on global monopoles in the AdS background. 
\section{Acknowledgment}
 We thank  Prof. Alberto Salvio for usefull suggestions.

\end{document}